\documentclass[showpacs,preprintnumbers,amsmath,amssymb,twocolumn,prb]{revtex4}

\usepackage{graphicx}

\usepackage{dcolumn}
\usepackage{bm}
\begin{document}

\preprint{APS/123-QED}

\title{In-plane field-induced vortex liquid correlations in 
underdoped Bi$_2$Sr$_2$CaCu$_2$O$_{8+\delta}$}

\author{Panayotis Spathis, Marcin Konczykowski, and Cornelis 
J. van der Beek}
\affiliation{Laboratoire des Solides Irradi\'{e}s, CNRS UMR 7642 \& 
CEA/DSM/DRECAM,  Ecole 
Polytechnique, 91128 Palaiseau, France}
\author{Piotr Gier{\l}owski}
\affiliation{\mbox{Institute of Physics of the Polish Academy of 
Sciences, 32/46 Aleja Lotnik\'ow, 02-668 Warsaw, Poland}}
\author{ Ming Li, Peter H. Kes}
\affiliation{\mbox{Kamerlingh Onnes Laboratorium, Rijksuniversiteit 
Leiden, P.O. Box 9506, 2300 RA Leiden, The Netherlands}}
\date{\today}

\begin{abstract}  The effect of a magnetic field component parallel to the 
superconducting layers on longitudinal Josephson plasma oscillations in 
the  layered high temperature superconductor Bi$_2$Sr$_2$CaCu$_2$O$_{8+\delta}$ is shown to depend  on 
the thermodynamic state of the underlying vortex lattice. Whereas the parallel 
magnetic field component  depresses the Josephson Plasma Resonance 
(JPR) frequency in the vortex solid phase, it may enhance it in the vortex liquid. 
There is a close correlation between the behavior of microwave absorption 
near the JPR frequency and the effectiveness of pancake 
vortex pinning, with the enhancement of the plasma resonance frequency 
occurring in the absence of pinning, at high temperature close to the vortex melting line.
An interpretation 
is proposed in terms of the attraction between pancake vortices and Josephson vortices, apparently also 
present in the vortex liquid state.

\end{abstract}
	
\pacs{74.20.De,74.25.Bt,74.25.Dw,74.25.Op,74.25.Qt,74.72.Hs}
\maketitle

\section{Introduction}

The application of magnetic fields oblique to the
superconducting planes of layered superconductors leads to 
novel vortex structures. Recent low--field scanning Hall probe 
\cite{grigorenkoN} and Lorentz microscopy experiments \cite{lorentz} 
on single crystalline Bi$_2$Sr$_2$CaCu$_2$O$_{8+\delta}$
have demonstrated the attraction between Josephson vortex (JV) and pancake vortex 
(PV) stacks, induced by, respectively, the field components $H^{\parallel}$ and 
$H^{\perp}$ parallel and perpendicular to the layers. This attraction leads 
to the formation of the so-called  combined lattice, 
\cite{Koshelev99} observed earlier in Bitter decoration.\cite{bolle}
In a ($H^{\parallel}$, $H^{\perp}$) phase diagram, the combined vortex 
lattice is delimited, as $H^{\perp}$ is increased, by a first order vortex 
melting transition to a vortex liquid at 
$H^{\perp}_{m}(H^{\parallel})$. \cite{Zeldov,Schmidt97,Koshelev99,Ooi99,Konczykowski2006}
As $H^{\parallel}$ is increased, there is another first order transition to a 
tilted PV lattice.\cite{Konczykowski2006} However, the actual structure of 
the different phases near the melting and combined-to-tilted transition lines remains speculative.

Whereas surface-sensitive techniques 
\cite{grigorenkoN,lorentz,Zeldov,bolle,Schmidt97,Koshelev99,Ooi99,Konczykowski2006} 
can establish phase boundaries, deeper insight in the structure of the vortex ensemble may be obtained from a bulk probe
such as Josephson Plasma Resonance (JPR).\cite{Matsuda95}  In analogy to JPR in a single junction, 
a microwave electric field applied perpendicularly to the superconducting 
layers of the Bi$_2$Sr$_2$CaCu$_2$O$_{8+\delta}$ compound (\em i.e. 
\rm parallel to the crystalline $c$-axis) 
induces a collective oscillation of Cooper pairs at the frequency 
$f_{pl} = ( s j_{c}^{c} / 2 \pi \Phi_{0} \varepsilon )^{1/2}$. Here, 
$j_{c}^{c}$ is the maximum $c$-axis (Josephson) critical current 
density, $\Phi_{0} = h/2e$ 
is the flux quantum, $s=1.5$ nm the interlayer spacing, and $\varepsilon$ 
the  dielectric constant. At the JPR frequency, the kinetic energy $h f_{pl}$ of the tunneling pairs is equal to the 
potential energy engendered by the time-periodic difference of the 
gauge-invariant superconducting order parameter phase  $\phi$  
between layers.\cite{Tachiki94,Koshelev1998} In the absence of a magnetic field component 
parallel to the superconducting layers, the JPR frequency squared is proportional to the 
time-- and disorder--averaged cosine of the difference of $\phi$ in adjacent layers $n$, 
$n+1$: \cite{Koshelev1998,colson} 
\begin{equation}
f^2_{pl}({\mathbf B},T)=f^2_{pl}(0,T) \langle \cos \phi_{n,n+1} \rangle \equiv f_{pl}^{2}(0,T) {\mathcal C}.   
\end{equation}
Since the spatial configuration of $\phi(\mathbf{r})$ is 
determined by the vortex distribution inside the sample, JPR is 
sensitive to the local structure of different vortex 
states.\cite{Koshelev1998} 

In the presence of a parallel magnetic field 
component, the $c$-axis electric field $E^{\perp}$ used to excite the JPR leads to periodic oscillations of the 
Josephson vortex lattice. Very recently, Koshelev explored a solution 
of coupled equations for plasma oscillations and JV lattice motion.\cite{Koshelev2007} 
In the limit of a dense JV lattice, $B^{\parallel} > B_{d} = \Phi_{0} / 
2 \pi\gamma s^{2}$, the frequency of longitudinal plasma oscillations 
with $q_{\perp} = 0$ (homogeneous displacement $u(\mathbf{r})$ of the JV 
lattice along the layers) is
\begin{equation}
f_{pl} = f_{pl}( H = 0 ) \sqrt{ 2 \langle \cos \phi_{0n} \rangle  + \frac{ 
B^{\parallel} }{ 2B_{d} } } \hspace{5mm} ( B^{\parallel} > B_{d} ).
\label{eq:in-plane-plasma}
\end{equation}
The JPR frequency increases with increasing parallel field, as the coupled 
system stiffens due to the increasing JV density.
In the above, $B^{\parallel} = \mu_{0}H^{\parallel}$, $\mu_{0} = 
4\pi\times 10^{-7}$~Hm$^{-1}$, $\gamma \equiv \lambda_{c}/\lambda_{ab}$ is the  
anisotropy of the $c$-axis and in-plane penetration depths, and 
\begin{equation}
\langle \cos \phi_{0n} \rangle 
                =    \frac{1}{2} \left[4 + \left( \frac{s}{  \lambda_{ab}}\right)^{2}\right] 
            \frac{ B_{d} }{ B^{\parallel} } \hspace{9mm} ( B^{\parallel} \gtrsim  2 B_{d} )
\end{equation}
is the average cosine of the phase 
\begin{equation}
 \phi_{0n}  = 2 \pi y / c_{y}  + \pi n + \alpha_{n}(y)    \hspace{15mm} ( B^{\parallel} > B_{d} )
\label{eq:inplanephase}
\end{equation}
in layer $n$. The phase consists of a monotonously increasing
contribution along the in-plane coordinate $y$, arising from the nearly homogeneous $H^{\parallel}$ 
component,  an in-plane oscillating contribution $\alpha_{n}(y)$, and 
an out-of-plane stepwise  contribution. The latter two terms reflect the structure of the JV lattice. The 
in-plane JV spacing is $c_{y} = \Phi_{0} / 2 \pi s B^{\parallel}$ in the dense 
limit, while $c_{y} = \sqrt{\beta  \gamma  \Phi_{0}/  B^{\parallel} }$ in 
the opposite, dilute limit ($\beta\equiv 2/\sqrt{3}$). Apart from the 
$q_{\perp}=0$ mode, a further, low-frequency mode  with $q_{\perp} = 
\pi / s $ and $f_{min} \approx 2 
f_{pl}( H = 0 ) \sqrt{ B_{d} / B^{\parallel} }$ is also 
predicted.\cite{Koshelev2007} This corresponds to antiphase motion of 
JV's in adjacent layers.

Experimentally, Josephson Plasma Resonance with a magnetic field applied obliquely 
to the superconducting planes of single crystalline 
Bi$_2$Sr$_2$CaCu$_2$O$_{8+\delta}$ has been studied by various authors.
\cite{Matsuda95,Tsui96,Matsuda97,Bayrakci99,Kakeya99} With the 
exception of Ref.~\onlinecite{Kakeya99}, these studies were limited to low temperatures $T < 0.6 T_{c}$ and 
$H^{\parallel} \gg H^{\perp}$, in the regime where the vortex liquid 
is the melt of the \em tilted  \rm PV lattice. The following results
have been obtained. In the vortex liquid, the perpendicular field component 
$H^{\perp}_{JPR}=H^{\perp}(f=f_{pl})$ at which 
the main JPR absorption peak occurs was found to decrease with increasing 
$H^{\parallel}$.\cite{Tsui96,Matsuda97,Bayrakci99} This is 
understood by the theory for a ``homogeneous'' vortex liquid without any interlayer PV 
correlations, developped in Ref.~\onlinecite{Koshelev1998}. Briefly, the 
short range  of the interlayer phase correlations inhibits the formation of a JV 
lattice, and leads to a very high effective anisotropy $\gamma$. Only 
the homogeneously increasing part of the  phase (\ref{eq:inplanephase}) 
remains; the JPR frequency decreases as the exponent of 
$H^{\parallel 2}$.\cite{Koshelev96,Matsuda97}
As the field is oriented closely to the superconducting 
layers,  a ``re-entrance'' of $H^{\perp}_{JPR}(H^{\parallel})$ occurs, such that $H^{\perp}_{JPR}$ now 
increases with $H^{\parallel}$.  This behavior was interpreted in 
terms of a model worked out by Bulaevskii {\em et al.} for the JV 
lattice in the dense limit, interacting with the PV 
lattice.\cite{Bulaevskii96,Bayrakci99}  The JPR frequency is thought to increase both 
as function of $H^{\perp}$ and $H^{\parallel}$ because of the larger 
JV lattice stiffness and the increasing JV-PV interaction (leading to a zig-zag PV 
lattice structure).\cite{Bulaevskii96} A caveat of this model is that, at fields 
($H^{\perp},H^{\parallel}$) at which the increase of the JPR frequency 
is observed in Bi$_2$Sr$_2$CaCu$_2$O$_{8+\delta}$, there is no 
evidence for a JV lattice.\cite{Konczykowski2006}

In this paper, we explore microwave dissipation in the regime 
of high temperature and moderate field angles, $H^{\perp} \lesssim H^{\parallel} 
\lesssim \mu_{0}^{-1}B_{d}$, in which the vortex liquid arises from 
the first order melting of the \em combined \rm lattice, that is, a 
vortex solid phase in which the  presence of JV stacks has been 
established.\cite{Schmidt97,Koshelev99,Ooi99,Konczykowski2006}
In this regime, we find a clear correlation between the behavior of $f_{pl}(H^{\parallel})$ and 
the thermodynamic state of the PV lattice as well as with PV pinning by material defects. In 
particular, we show that while increasing $H^{\parallel}$ in the vortex solid phase 
yields a modest decrease of $f_{pl}$, the JPR frequency in the vortex liquid phase close to the melting line 
is \em enhanced \rm by the additional in-plane field. This is in 
agreement with data of Refs.~\onlinecite{Matsuda97}, \onlinecite{Kakeya99} and 
\onlinecite{Morozov99} that suggest an increase of both $f_{pl}$
and the $c$-axis conductivity with increasing $H^{\parallel}$ under 
similar experimental conditions. We conclude that in the regime under 
study the vortex liquid has a correlated 
character due to the persistence of Josephson vortices. Moreover, the sharp dependence 
of $f_{pl}$ on PV density in the liquid reveals a boundary between this correlated  liquid and 
the homogeneous vortex liquid at higher fields.

\begin{figure}[t]
\begin{center}
\includegraphics[width=7.8cm,keepaspectratio]{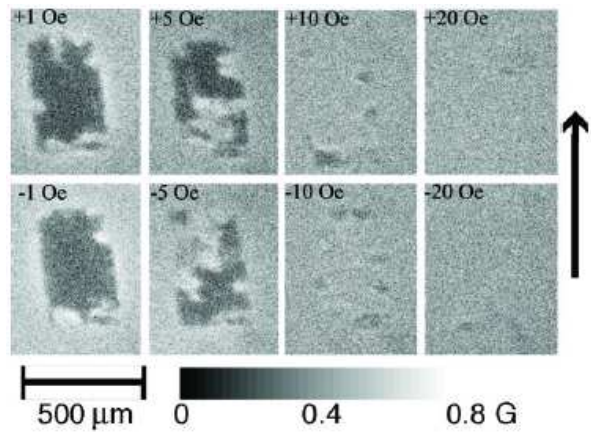}
\caption{\label{fig:magneto}Differential magneto-optical images of 
the underdoped Bi$_{2}$Sr$_{2}$CaCu$_{2}$O$_{8+\delta}$ single 
crystal ($0.28\times0.46\times0.04$ mm$^3$) at $T=60$ K. An in-plane 
field $H^{\parallel} = 500$ Oe was applied in the direction of the 
arrow. Images were obtained by averaging 10 CCD frames taken at the 
field $H^{\perp} + \Delta H^{\perp}$ and subtracting the average of 
10 frames taken at $H^{\perp}$. This procedure was repeated 100 
times. The field modulation $\Delta H^{\perp}$ was 0.5 Oe. The intensity 
scale shows the magnitude of the ac component of the local induction, 
$\Delta B^{\perp} = B^{\perp}(H^{\perp}+\Delta 
H^{\perp})-B^{\perp}(H^{\perp})$.
} \vspace{-5mm}
\end{center}
\end{figure}

\section{Experimental details}

Single crystals ($T_c=72.4 \pm 0.6$ K) were cut from a batch of 
underdoped Bi$_2$Sr$_2$CaCu$_2$O$_{8+\delta}$,\cite{li} and characterized
using the differential magneto-optical technique (DMO). \cite{soibel} 
By taking the difference between the average of 10 consecutive 
magneto-optical images acquired at perpendicular fields 
$H^{\perp}+\Delta H^{\perp}$ and $H^{\perp}$, one obtains the local 
magnetic permeability at field modulation $\Delta H^{\perp}$ 
(Fig.~\ref{fig:magneto}). Underdoped 
Bi$_2$Sr$_2$CaCu$_2$O$_{8+\delta}$ being prone to significant disorder 
in oxygen distribution,\cite{vdBeek2003} the DMO technique was used 
to select the most homogeneous crystals for further experiments. 
Nevertheless, some inhomogeneity in the penetration of perpendicular 
magnetic flux remains, as shown in Fig.~\ref{fig:magneto}. 
A clear dependence of the direction of flux penetration was 
observed, depending on the relative orientation of $H^{\perp}$ and 
$H^{\parallel}$. The DMO technique was also used to detect the 
irreversibility line $B_{irr}(T)$, at which the permeability becomes 
unmeasureably small due to the demise of vortex pinning, both with $H^{\parallel}=0$ and 
$H^{\parallel}=500$ Oe. 

\begin{figure}[t]
\begin{center}
\includegraphics[width=8.7cm,keepaspectratio]{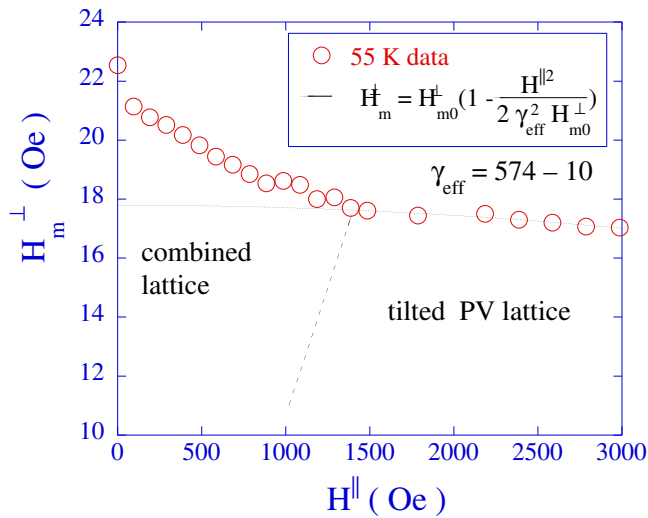}
\caption{ \label{fig:Parallel-melting} First order vortex lattice melting 
field of the underdoped Bi$_2$Sr$_2$CaCu$_2$O$_{8+\delta}$ crystal of 
Fig.~\protect\ref{fig:magneto}, at $T = 55$ K. As the in-plane field 
$H^{\parallel}$ increases, $H_{m}^{\perp}$ first decreases linearly. 
At high in-plane fields, the parabolic behavior
$H_{m}^{\perp} \approx H_{m0}^{\perp}\left( 1 - H^{\parallel 2} / 2 
\gamma_{eff}^{2}H_{m0}^{\perp 2} \right)$ correpsonding to the 
anisotropic London model is followed.\protect\cite{Konczykowski2006}  The dashed line indicates 
the boundary between comined lattice and titled pancake vortex lattice 
phases.  }
\vspace{-8mm} 
\end{center} 
\end{figure}

The local magnetization was measured using a microscopic Hall probe array.\cite{Zeldov,Konczykowski2006} 
The first order melting field $H^{\perp}_{m}$ of the vortex lattice was 
determined from the sharp paramagnetic peak that melting induces in the first and third harmonic of the 
transmittivity.\cite{Morozov96,Schmidt97,Konczykowski2006} 
The temperature dependence of the melting field in $H^{\parallel} =  0$ is well described by 
$H_{m}^{\perp} \sim A \Phi_{0}/\mu_{0}\gamma s \lambda_{ab} (\varepsilon_{0}s/k_{B}T)$, where 
$\varepsilon_0=\Phi_0^{2}/4\pi\mu_{0}\lambda_{ab}^2$, and $A \approx 1$.\cite{Blatter96} 
The dependence of $H_{m}^{\perp}$ on $H^{\parallel}$ is in every way similar to that 
observed in optimally doped single crystalline Bi$_2$Sr$_2$CaCu$_2$O$_{8+\delta}$ (see 
Fig.~\ref{fig:Parallel-melting}). The melting field of the combined lattice 
follows a linear $H^{\parallel}$-dependence, while the melting of the 
tilted PV lattice at high $H^{\parallel}$ follows a parabolic 
dependence. From a fit of the latter to the anisotropic London model,\cite{Konczykowski2006} we obtain an 
effective anisotropy parameter $\gamma^{eff} \approx 574\pm 10$, close to 
the bare value $\gamma \approx 600$ obtained previously for underdoped  
Bi$_2$Sr$_2$CaCu$_2$O$_{8+\delta}$.\cite{colson}

The microwave dissipation was measured using the cavity perturbation technique.\cite{colson}
Two different oxygen-free high-conductivity copper resonant cavities, 
with the $TM_{010}$ mode at 19.2 and 38.7 GHz respectively, were used in the measurements. 
The sample was glued in the center of the cavity end-plate so that 
the  microwave electrical field $E^{\perp}$ was perpendicular to the 
superconducting layers  when the cavity was operated in the TM$_{01i}$ 
resonant modes. For cavity I, the unloaded quality factors $Q$ of five  modes with $i=0,...,4$ 
were measured as a function of  magnetic field and temperature, allowing us to probe the microwave dissipation  
in both the vortex liquid and vortex solid state during the same run. 
For cavity II, only the $TM_{010}$ and $TM_{011}$ modes were used.
For all modes, the sample lateral dimensions ( typically $< 500$ $\mu$m ) 
were sufficiently small with respect to the cavity size for 
$E^{\perp}$ to be considered homogeneous, and the microwave magnetic field to be 
negligible. This was checked by cutting the most important crystals in 
two, and remeasuring the microwave absorption ({\em e.g.}, on the 
crystal of Fig.~\ref{fig:magneto}).

Two orthogonal coils were used to control $H^{\parallel}$ and $H^{\perp}$ independently. 
Measurements were carried out by first applying 
$H^{\parallel}$ in the normal state, by setting the desired 
temperature, and then sweeping $H^{\perp}$. The misalignment of 
$H^{\perp}$ did not exceed $1^{\circ}$, entailing a variation of $H^{\parallel}$ 
of less than $2$ Oe. The symmetry of the microwave response with respect to $H^{\perp}$ and 
the absence of parasitic field components was checked by 
sweeping $H^{\perp}$ up to the maximum positive value, from there down 
to the maximum negative value and back to zero.

\begin{figure}[t]
\begin{center}
\includegraphics[width=8cm,keepaspectratio]{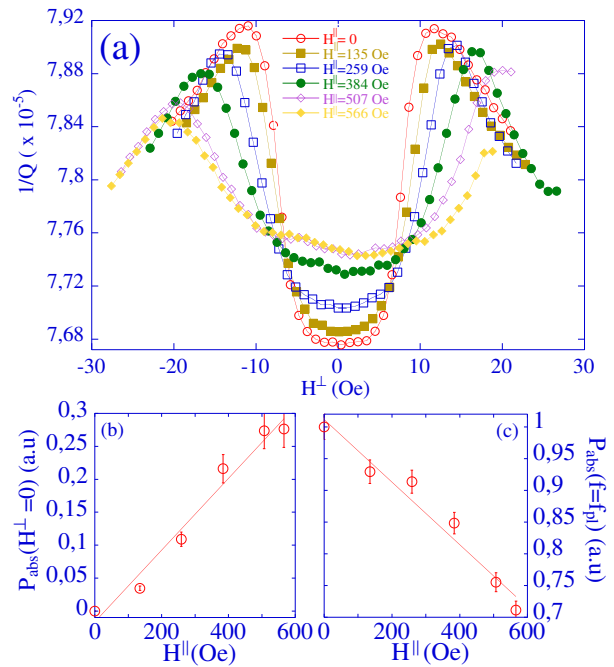}
\caption{ \label{fig:microwave}(a) Magnetic-field dependence of the 
microwave absorption  obtained during a sweep of the perpendicular 
field, for different in-plane fields at $66$ K and $19.2$ GHz. All 
data were taken during the same run, allowing for a quantitative 
comparison of the power absorption. The temperature was raised above 
$T_c$ between runs with different in-plane fields. (b) Microwave 
losses for $H^{\perp}\approx 0$: 
$P_{abs}(H^{\perp}=0)=[Q^{-1}(0,H^\parallel)-Q^{-1}(0,0)]/[Q^{-1}(H^{\perp}_{JPR},0)-Q^{-1}(0,0)]$. 
(c) Magnitude of the absorption at the JPR: 
$P_{abs}=[Q^{-1}(H^{\perp}_{JPR},H^\parallel)-Q^{-1}(0,0)]/[Q^{-1}(H^{\perp}_{JPR},0)-Q^{-1}(0,0)]$.}

\end{center} \vspace{-5mm}
\end{figure}

\section{Results}

Fig.~\ref{fig:microwave}(a) shows the microwave absorption
 at 66 K and 19.2 GHz during the $H^{\perp}$-sweeps, 
for different values of $H^{\parallel}$. For 
$H^{\perp}\approx 0$, the power absorption linearly increases with 
in-plane field [Fig.~\ref{fig:microwave}(b)]. Upon increasing 
$H^{\perp}$ from zero, the absorption rises, and a clear maximum of 
$Q^{-1}$ appears, corresponding to the excitation of the Josephson Plasma 
Resonance at $f=f_{pl}(H,T)$.\cite{Matsuda95,Tachiki94}  The independence 
of the shape and position of the measured JPR peaks on sample size confirm that  the 
longitudinal JPR mode is observed. Performing the measurement for non-zero 
parallel field component, one sees that the maximum of the 
absorption shifts to higher $H^{\perp}$ as $H^{\parallel}$ increases. 
The low-$H^{\perp}$ flank of the absorption peak is progressively 
suppressed with $H^{\parallel}$, while the high-field tail is nearly 
$H^{\parallel}$--independent; concordingly, the absorption peak 
progressively diminishes in height [Fig.~\ref{fig:microwave}(c)].

\begin{figure}[t]
\begin{center}
\includegraphics[width=8.7cm,keepaspectratio]{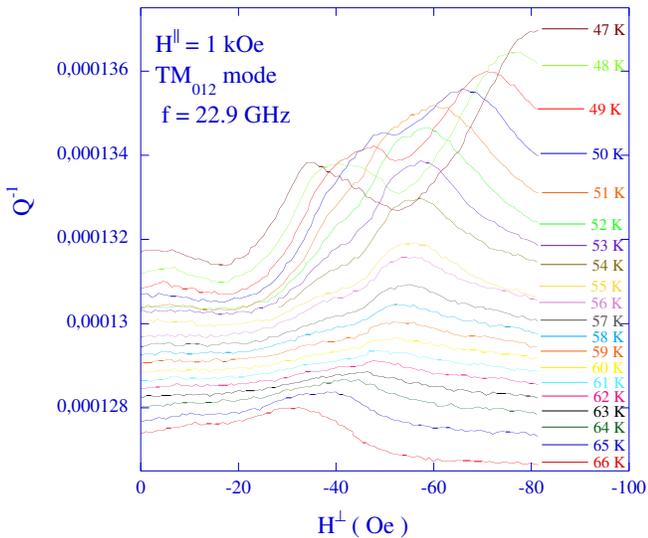}
\caption{ \label{fig:TM012-H||1kOe-Tdep}(a) Dependence on 
perpendicular field $H^{\perp}$ of the microwave absorption at $22.9$ GHz, 
at different temperatures. The parallel field $H^{\parallel} = 1$ 
kOe. Data are plotted as one over the unloaded quality factor of the 
resonant cavity. At high temperatures, 
there is a single absorption maximum. At $T \lesssim 55$ K, a second, 
low $H^{\perp}$--peak appears. All field sweeps have 
been performed during a single run, with the sample being heated above 
$T_c$ between different measurement temperatures.  For clarity, successive curves have been displaced 
by $\protect\Delta Q = 30$.}

\end{center} \vspace{0mm}
\end{figure}

The temperature evolution of the absorption during $H^{\perp}$--sweeps performed with 
an in-plane field $H^{\parallel} = 1$ kOe is depicted in Fig.~\ref{fig:TM012-H||1kOe-Tdep}. 
For temperatures above 61 K, a single absorption maximum is observed, and the behavior at all 
investigated in-plane fields is that represented by 
Fig.~\ref{fig:microwave}(a). This peak corresponds to the ``high 
temperature resonance'' reported in Refs.~\onlinecite{Kakeya99,Kakeya2005}. 
Its height decreases as function of $H^{\parallel}$ and temperature. For $T \lesssim 61$ K and 
in-plane fields in excess of 600 Oe, a shoulder appears, that 
develops into a second, low-field absorption maximum as the temperature 
is lowered. The two low-temperature absorption peaks correspond to the 
reentrant  ``low temperature resonance'' of Refs.~\onlinecite{Kakeya99,Kakeya2005}. 
The magnitude of microwave absorption on the high--field branch of 
this ``low-temperature resonance'' is very weakly dependent on $H^{\parallel}$ and nearly equal 
to that measured in $H^{\parallel} = 0$. Unlike previous reports,\cite{Kakeya99,Kakeya2005}
clear peaks in the absorption could not be observed when the magnetic field was perfectly
 aligned with the superconducting layers. 

\begin{figure}[t!]
\begin{center}
\includegraphics[width=8cm,keepaspectratio]{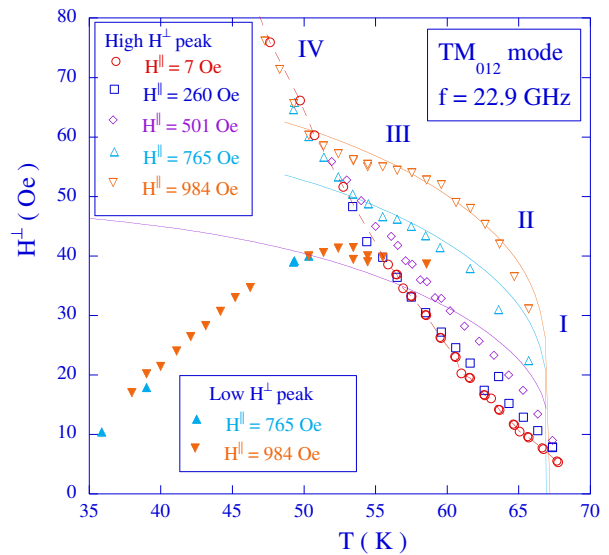}
\caption{\label{fig:TM012-B(T)-with-fit} Temperature dependence of JPR fields 
obtained for $22.9$ GHz for various in-plane fields. Open symbols 
represent the dissipation peak measured at high perpendicular field, 
closed symbols represent the position of the low-$H^{\perp}$ maximum 
observed for $H^{\parallel} > 500$ Oe.  Drawn lines represent 
the boundary (\ref{eq:boundary}), evaluated with $\tilde{C} =2.6$, at 
which phase correlations in the liquid have dropped back to their 
value in zero in-plane field. The dashed line is a guide to the eye 
representing the $H_{JPR}^{\perp}$-locus for $H^{\parallel}=0$.}
\label{fig:TM012-B(T)-diagram-with-fit}
\end{center} \vspace{-5mm}
\end{figure}

\begin{figure}[t!]
\begin{center}
\includegraphics[width=8.5cm,keepaspectratio]{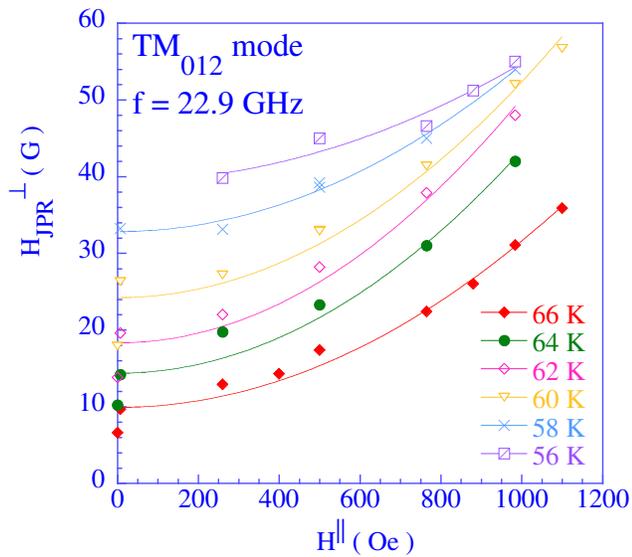}
\caption{ \label{fig:B||} In-plane field-dependence of $H^\perp_{JPR}$ 
at  $22.9$ GHz.}
\end{center} \vspace{-5mm}
\end{figure}

Fig.~\ref{fig:TM012-B(T)-diagram-with-fit} collects the perpendicular field component  
$H^{\perp}_{JPR}\equiv H^\perp(f_{pl} = 22.9$ GHz) at which the JPR absorption 
maximum is located, for different magnitudes of the in-plane field. 
For this particular frequency, the main JPR peak occurs in the vortex liquid phase. 
Comparing the evolution of the JPR for different values of 
$H^{\parallel}$ to the case where $H^{\parallel} \approx 0$, one can distinguish 
four different temperature regimes. Within 2 K of $T_c$, microwave absorption is very 
large and the resonant field cannot be identified (regime I). As the temperature is lowered, %
the JPR becomes observable. The JPR field sharply increases with decreasing $T$ (regime 
II), and then slows down to a plateau (regime III). In this regime, 
$B_{JPR}^{\perp}$ increases nearly quadratically with $H^{\parallel}$ 
(Fig.~\ref{fig:B||}). We note that the upward shift of the lines 
of constant $f_{pl}$ in regimes II and  III implies the 
\em increase of $f_{pl}$  itself \rm when an in-plane field is applied and 
$H^{\perp}$ is kept constant. Finally, at low $T$, 
$H^\perp_{JPR}$ is smaller than it is for $H^{\parallel} = 0$ (regime IV). The 
observed decrease in this regime is larger for lower frequencies, 
\textit{i.e.} higher temperatures, and is described by the 
high-temperature expansion.\cite{Koshelev96,Matsuda97,Koshelev1998}

In Fig.~\ref{fig:JPR}, we compare the position of $B^{\perp}_{JPR}$ at three different frequencies
to the main features of the vortex phase diagram. In the absence of an in-plane field [Fig.~\ref{fig:JPR}(a)],
the JPR at $19.2$ and $22.9$ GHz occurs in the vortex liquid state. The 
application of a parallel field component leads to the increase of 
$H_{JPR}^{\perp}$ and the four regimes described above. However, when the JPR occurs in 
the vortex solid, $H^{\perp}_{JPR}$ shows a moderate decrease when an in-plane field 
is applied [Fig.~\ref{fig:JPR}(b)].

\section{Discussion}

\begin{figure}[t!]
\begin{center}
\includegraphics[width=8.3cm,keepaspectratio]{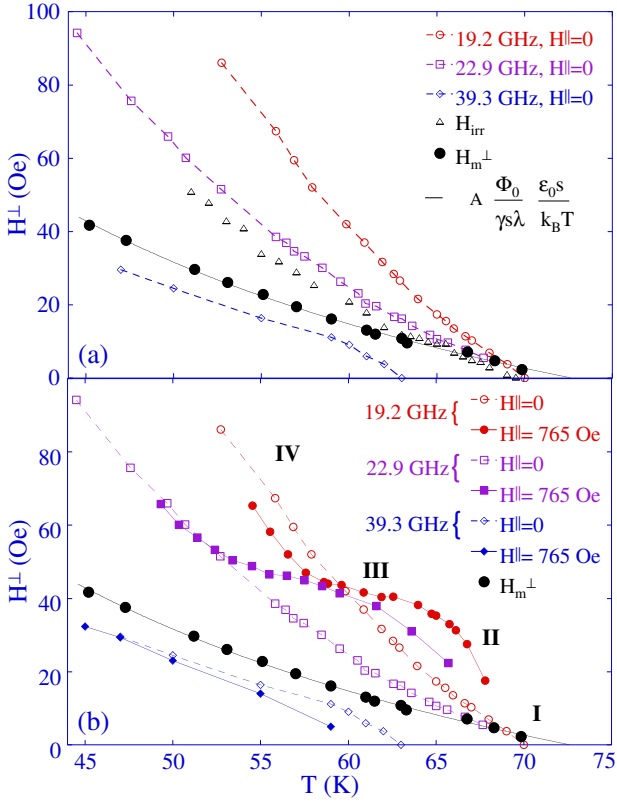}
\caption{ \label{fig:JPR} (a) $T$-dependence of $H^\perp_{JPR}$ 
obtained for $19.2$, $22.9$ and $39.3$ GHz. The irreversibility line 
and the vortex lattice melting line for $H^{\perp}=0$ are also 
reported. The fit of the melting line $H_{m}^{\perp} = A (\Phi_0/\lambda_{J} 
\lambda)(\varepsilon_0 s/k_B T)$ is taken from 
Ref.~\cite{Blatter96} ($A \approx 1$) (b) 
Effect of $H^{\parallel}$ on the JPR fields. For clarity, only results 
obtained for one value of the in-plane field for each frequency are 
shown. Labels I to IV refer to regimes described in the text and 
apply to the 19.2 and 22.9 GHz curves.}
\end{center} \vspace{-5mm}
\end{figure}

The behavior of the field at which the JPR is excited can be summarized as 
follows. In the vortex solid phase, and in the vortex liquid at low 
temperature [regime IV, in Figs.~\ref{fig:TM012-H||1kOe-Tdep} and \ref{fig:JPR}(a)], 
$H_{JPR}^{\perp}$ decreases as $H^{\parallel}$ is 
increased. These regimes correspond to those parts of the 
$(H,T)$--phase diagram in which pinning of the pancake vortices due to 
material defects is strong. Notably, in underdoped 
Bi$_{2}$Sr$_{2}$CaCu$_{2}$O$_{8}$, which is characterized by the 
presence of numerous oxygen vacancies in the superconducting 
CuO$_{2}$-planes,\cite{Li96ii} as well as substantial disorder in the 
oxygen distribution, one has strong pancake vortex pinning in the vortex solid phase.\cite{vdBeek2003}
On the other hand, the featureless DMO images above the first order melting line 
show that vortex pinning in the vortex liquid at high temperature is absent. 
Then, $H_{JPR}^{\perp}$ dramatically increases with $H^{\parallel}$, signalling an 
increase of the Josephson plasma frequency.\cite{note-on-pinning} In regime I, for $T > 67$ K, no features 
related to the JPR can be observed when a parallel field component is applied.

The above observations can be consistently explained by invoking the 
presence of Josephson vortices in both the vortex solid (combined 
lattice) \em and \rm the vortex liquid phases, and their interaction 
with the microwave  electrical field $E^{\perp}$ and with pancake 
vortices. The interaction of JV's with  $E^{\perp}$ only is 
brought out by the absorption at $B^{\perp} = 0$. In our experimental configuration, the  
$E^{\perp}$ induces a $c$-axis current that drives the JV's, thus 
producing flux-flow losses.\cite{bontemps,Matsuda97,Kakeya2005} 
The dissipated microwave power is proportional to $E^{\perp 2}/\rho_{f} \propto 
f^2 H_{\parallel}^2 /\rho_{f}$, where $\rho_{f}$ is the JV flux-flow 
resistivity. Experimentally, the dissipated power $ \sim Q^{-1}$ is 
found to be linear in $H^{\parallel}$ [Fig.~\ref{fig:microwave}(b)]. This implies the linearity of  
$\rho_{f}$ with in-plane field, in full agreement with the prediction of Koshelev 
\cite{Koshelev1998} and the experiments of Latyshev \cite{latyshev} 
in the dilute limit of the JV lattice, $B^{\parallel} < B_{d}$. 

The effect of the PV lattice on JV motion was recently explored by 
Koshelev, Latyshev, and Konczykowski,\cite{latyshev2006} who found 
that the attractive interaction between a moderate density of pancakes 
and the JV lattice induces significant damping of the JV motion and 
an increase of $j_{c}^{c}$ in overdoped 
Bi$_{2}$Sr$_{2}$CaCu$_{2}$O$_{8}$ whiskers. Moreover, heavy-ion irradiation of the 
whiskers introduces  defects that strongly pin the PV's, 
overwhelming the PV-JV attraction and re-establishing the JV 
mobility. We surmise that in our experiment, PV's in the vortex solid are strongly 
pinned; therefore, the microwave electric field drives large-amplitude JV 
oscillations. We then expect the JPR frequency to be described by 
Eq.~(\ref{eq:in-plane-plasma}). The main effect of the PV lattice is 
a small random contribution to $\phi_{0n}$ arising from thermal 
fluctuations, and a concomitant reduction of $\langle \cos \phi_{0n} \rangle$ 
as a function of $H^{\perp}$. This entails the modest decrease of 
$H_{JPR}^{\perp}( H^{\parallel} )$ that is experimentally observed.

We now turn to regimes II and III, in which $H_{JPR}^{\perp}$ increases 
quadratically as function of $H^{\parallel}$. The increase of the 
characteristic frequency of this high temperature plasma mode was 
interpreted by Kakeya \em et al. \rm as being due to the plasma wave 
propagation at $\mathbf{q} = ( q_{\parallel} = 2\pi/c_{y}, 
q_{\perp} = \pi/s)$. The dispersion relation of the longitudinal plasma, $f_{pl} = 
f_{pl}(\mathbf{q}=0) \sqrt{ 1 + \lambda_{ab}^{2}q_{\perp}^{2}}$, would 
then be responsible for the observed $H^{\parallel}$ dependence. However, the 
dispersion of the longitudinal plasma is too weak to explain the 
experimentally measured variation, a discrepancy that is worse if 
charging effects are taken into account.\cite{Tachiki94} Moreover, the motion of JV's 
in this particular temperature-field regime is expected to be strongly 
damped due to the interaction with (unpinned) pancake vortices. We therefore seek an 
interpretation in terms of nearly immobile JV's. The increase of 
the JPR frequency must then be due to the enhancement of $c$-axis 
phase correlations by the in-plane field.\cite{kosh5} 
                                
Phase correlations in the vortex liquid can be 
increased either by macroscopic segregation into PV-rich and 
PV-depleted zones, or alternatively, by local rearrangements of PVs 
to adjust to the presence of JVs, even in the 
vortex liquid  state. As to phase segregation, recent 
magneto-optical imaging on overdoped Bi$_2$Sr$_2$CaCu$_2$O$_{8+\delta}$ for 
fields very closely aligned to the planes has shown that large inhomogeneities 
of the PV density can occur under the influence of an in-plane field.\cite{tam} 
We have performed the same experiment 
to verify that this is not the cause of the change of the 
JPR response. For low values of $H^{\perp}$, the field penetration is indeed 
inhomogeneous (Fig.~\ref{fig:magneto}). In particular, as in 
Ref.~\onlinecite{yak}, 
the shape of the phase transformation front at first order vortex melting is
 affected by the direction of the in-plane field.  The two field components 
are therefore not independent within the sample. However, once the vortex 
system is entirely transformed to the liquid phase, no contrast can be seen 
using the DMO technique. This means that inhomogeneity of $B^{\perp}$ 
is less than our resolution ($0.1$ Oe) and cannot explain the changes of the JPR 
when an in-plane field is applied.

We therefore propose that PVs in the vortex liquid  rearrange 
themselves on short length scales in order to coincide with JVs, yielding a 
novel correlated, PV density-modulated liquid. According to Bulaevskii \cite{Bulaevskii96} and 
Koshelev,\cite{Koshelev99} the Lorentz force exerted by the in-plane 
currents of JVs on a PV is balanced by the restoring force of the 
other pancakes. In the vortex solid, this balance results in a mutual 
attraction between a JV and a flux line, and the formation of the 
crossing-lattice structure.\cite{grigorenkoN,lorentz} The range of 
this attractive interaction is of the order of $\lambda_J=\gamma 
s$. The energy gain per unit length due to a crossing event between a 
PV stack and a JV is $\displaystyle E_{\times}=4 C_{solid}  
s \varepsilon_0^2 /\lambda_J^2 U_M$. Here, $U_M=C_{44}k_z^2/2 n_v$ is 
the magnetic tilt stiffness of the pancake vortex stack, 
$C_{44}=B^\perp\varepsilon_0\ln\left[1+r_{cut}^2 k_z^2/(1+r_w^2 
k_z^2)\right]/ 2\Phi_0\lambda_{ab}^2 k_z^2$,\cite{koshkes} 
$n_v=a_0^{-2}=B^\perp/\Phi_0$, $k_z\approx \pi/s$ is the typical 
wavevector of the pancake stack deformation, 
$r_{cut}=\min(a_0,\lambda_{ab})\approx \lambda_{ab}$, and $C_{solid}$ 
 a constant. The wandering length $r_w$ describes the relative 
displacement of PVs in neighboring layers.\cite{koshC}
Carrying over the calculation to the vortex liquid, we find that 
$E_{\times}$ is particularly well estimated by these expressions. 
Namely, these disregard the Josephson coupling term in the vortex line 
tension, absent in the liquid.\cite{colson} However, $r_w$ 
is not defined in the vortex liquid and should be replaced by 
$\sqrt{2\Phi_0(1-{\cal C})/\pi B^\perp}$.\cite{koshC} 
Also, we replace $C_{solid}$ by $C_{liquid}$.

A comparison of the energy $E_{\times}/c_z$ gained from crossing 
events to the PV interaction energy $E_{int}=C_1 
\varepsilon_0 K_0(a_0/\lambda_{ab})$ shows that rearrangement of PVs in the 
vortex liquid so as to maximize the number of  crossing events with JVs is favorable if
\begin{eqnarray}
8\sqrt{\frac{\gamma B^\parallel}{\beta 
\Phi_0}}\left(\frac{\lambda_{ab}^2}{\gamma^2 
s}\right)\Biggl/\ln\left(1+\frac{\lambda_{ab}^2\pi B^\perp}{2\Phi_0 
(1-{\cal C})}\right)>\tilde{C}K_0\left(\frac{a_0}{\lambda_{ab}}\right)
\label{eq:boundary}
\end{eqnarray}
Here, $K_0(x)$ is the modified Bessel function, $c_z$ is the JV lattice (JVL) 
spacing along the $c$-axis,\cite{camp} and  $\tilde{C}=C_1/C_{liquid}$. Even 
though the inequality (\ref{eq:boundary}) is implicit
because of the dependence of ${\cal C}$ on the crossing energies, it can be  
evaluated using the experimental values of ${\cal C}$ 
at a given measuring frequency. 

Solving Eq.~(\ref{eq:boundary}) for $B^{\perp}(T)$, we obtain 
correlated-homogeneous vortex liquid boundaries for different values of 
the in-plane field, see Fig.~\ref{fig:TM012-B(T)-diagram-with-fit}. 
Here, we have taken $\lambda_{ab}(T)=\lambda_{ab}(0)/\sqrt{1-(T/T_{c})^4}$, with 
$\lambda_{ab}(0)=300$ nm.\cite{colson} For $\tilde{C}=2.6$, these boundaries coincide with the loci of 
$f_{pl}$ in regimes II and III, which in turn correspond to the lines on which 
${\mathcal C}$ 
for non-zero $H^{\parallel}$ has dropped back to the value 
observed near the melting line in the \rm absence \rm of an in-plane field.
Thus, the observed increase of $H_{JPR}^{\perp}(T)$, and consequently, of the interlayer 
phase correlations, is fully consistent with the existence of a 
correlated vortex liquid in the region delimited by the dashed and 
drawn lines of  Fig.~\ref{fig:TM012-B(T)-diagram-with-fit}, the extent of which grows 
with increasing $H^{\parallel}$. In the correlated liquid, the 
attraction between PVs and JVs leads to a spatially modulated PV 
density: in the vicinity of JV stacks, the mean distance between PVs
is smaller than in JV-free regions. 
The $c$-axis phase coherence of this state is higher than that of the vortex liquid for 
$H^{\parallel}=0$ because of the depletion of PVs in the regions between 
JV stacks. The demise of the correlated liquid at high $T$ 
(regime II) is caused by the  decrease with temperature of 
the inter-pancake vortex repulsion, less rapid than that of the PV-JV attraction.
The disappearance of the correlated liquid  when  $H^{\perp}$ is 
increased is very sudden, as testified by the close proximity of 
the resonant fields for $f=19.2$ and $22.9$ GHz in regime 
III [Fig.~\ref{fig:JPR}(b)]. This shows that the phase coherence in the 
plateau-regime is nearly constant, and then rapidly drops as the PV 
density is increased.

The modification of the lineshape of the JPR absorption supports our 
interpretation. Fig.~\ref{fig:microwave}(c) shows that the maximum absorption 
at $f=f_{pl}$ in regime II decreases linearly as function of 
$H^{\parallel}$. This is explained by the presence of well-defined JVs. 
Namely, the plasma oscillation is impeded in the vicinity of 
the non-linear JV cores because of the rapid $2\pi$-modulation, with period 
$c_z$, of the gauge-invariant phase $\phi$ along the $c$-axis. 
The resulting decrease in dissipation is given by 
the ratio $\lambda_J s H^\|/\Phi_0$ of the volume occupied by the JVL 
(\textit{i.e.} that does not participate in the JPR), to the volume 
of the sample. Cooper pairs located in JV-free regions still exhibit 
plasma oscillations as long as the separation $c_{y}$ between JVs is larger 
than $L_{\phi}={\lambda_J}^2/a_0$.\cite{kosh5} This is well satisfied: 
for our in-plane field range, $260 < H^{\parallel} < 1100$ Oe, $c_y\approx(5-10) L_{\phi}$. 
It should be noted that the coefficient of the linear decrease of absorption is ten times higher than 
expected. This is, possibly, because the width of JV-core region affecting the JPR is 
larger than $\lambda_J$.

The measurements also show that the low-field flank of the absorption 
peak decreases with $H^{\parallel}$, whereas the high field tail is 
only pushed slightly upwards. In the vortex liquid state, the low 
field part of the JPR absorption comes from localized JPR modes over 
wide areas of the sample.\cite{kosh5} However, the presence of a JVL 
prevents the occurrence of homogeneous Josephson coupling over such 
areas. The high field part of the absorption arises from 
contributions of small fluctuating areas over which Josephson 
coupling is established in a hap-hazard way. These local 
contributions are not destroyed by the JVL.

Finally, we address the low temperature region IV, in which the main JPR 
absorption peak lies above the postulated correlated-homogeneous 
vortex liquid boundary. From the height of the absorption peak, which 
is nearly $H^{\parallel}$--independent and close to what is observed in 
$H^{\parallel} = 0$, we conclude that the whole sample volume 
contributes to the plasma resonance. This is consistent with the 
absence of Josephson vortices, the formation of which is inhibited by
pancake vortex fluctuations. The JPR is best described by the high temperature 
expansion of Refs.~\onlinecite{Koshelev96,kosh5}. 

On the other hand, the low $H^{\perp}$ peak, which corresponds to the 
``low-temperature mode'' of Refs.~\onlinecite{Kakeya99,Kakeya2005}, 
occurs only in the vortex solid and in the correlated vortex liquid. 
It is therefore related to the existence of well-defined JV's. We note 
that the very weak $H^{\parallel}$--dependence of this peak seems to
contradict the $H^{\perp}$--independent, $\sim H^{\parallel -2}$ dispersion 
of the predicted antiphase mode with $q_{\perp} = \pi / 
s$.\cite{Koshelev2007} 

\section{Summary and conclusion}

In summary, the effect of an in-plane magnetic field on microwave 
dissipation near the longitudinal Josephson 
Plasma Resonance in underdoped Bi$_2$Sr$_2$CaCu$_2$O$_{8+\delta}$ is 
closely correlated to the thermodynamical state of the underlying 
vortex system, and, in particular, to the effectiveness of pancake 
vortex pinning. In the vortex solid, the application of an in-plane 
field slightly suppresses the perpendicular field at which the JPR 
takes place. In contrast, the application of an in-plane field in the
presence of perpendicular fields exceeding the first order melting transition 
can lead to a correlated, PV density-modulated vortex liquid. The overall 
phase coherence of this phase is higher than the coherence of the usual vortex liquid state because 
a substantial fraction of pancakes is aligned on stacks of Josephson 
vortices. The existence of well-defined Josephson vortices, and 
therefore, of a correlated liquid, requires 
the integrity of PV stacks over, at least, the distance $c_{z}$ 
separating Josephson vortices in the same stack along the $c$-axis; 
hence, it is only possible for low densities of PVs, close to the 
melting line. Our results therefore constitute evidence for the ``linelike nature'' of 
the vortex liquid close to the melting line in 
Bi$_{2}$Sr$_{2}$CaCu$_{2}$O$_{8}$, similar to the ``linelike liquid'' 
reported in Ref.~\onlinecite{thorsmolle} for less anisotropic layered
Tl$_{2}$Ba$_{2}$CaCu$_{2}$O$_{8}$. The correlated-homogeneous vortex 
liquid boundary is also reminiscent of recent results showing 
a transition from a linelike liquid to a phase without vortex 
line tension in YBa$_{2}$Cu$_{3}$O$_{7-\delta}$.\cite{bouquet}

\section{Acknowledgements}
We thank J. Blatter, A.E. Koshelev,  D. van der Marel, and T. Shibauchi  
for useful discussions. One of us (P.G.) was partially supported by MNiSW-grant no. N202 058 32/1202.

\bibliography{BIBLIOGRAPHY}

\begin{thebibliography}{27}
\bibitem{note-on-pinning}
  \bibinfo{author}{\bibfnamefont{The observed increase of 
  $H_{JPR}^{\perp}$  with $H^{\parallel}$ is much weaker or absent in more disordered 
  crystals in which the irreversibility line lies above the measured 
  Josephson Plasma Resonance peaks.}}
\end{thebibliography}
\bibliographystyle{apsrev}

\end{document}